\documentclass{article}  
\title{{\bf{On the Hopf Structure  of \\ $W_2$ - Algebra and
        N=1 Superconformal Algebra in the OPE Language}}}
\author{\bf{H.\ T.\"Ozer}\thanks{\bf{e-mail\ :\ ozert @ itu.edu.tr}} \\\\
Physics Department,\ Faculty of Science and Letters,\\
Istanbul Technical University,\\
80626,\ Maslak,\ Istanbul,\\
Turkey
}

\begin{document}      
\maketitle 
\begin{abstract}                
Hopf structure of the prototype realizations of the $W_2$-algebra and also
$N=1$ superconformal algebra are obtained using the bosonic and  also fermionic 
Feigin-Fuchs type of free massless scalar fields in the operator product 
expansion (OPE) language.
\end{abstract}

\setcounter{equation}{0}

\vskip 5mm
\section{\bf{Introduction}}\label{ch:int}
\vskip 3mm
\noindent Conformal symmetry has played an important role in the developments 
of the physics contents of such models: i.e. (super)string theory \cite{1},statistical
physics \cite{2}, as well as in mathematics \cite{3}. Its underlying symmetry 
algebra is Virasoro algebra which is a Lie algebra.  It is well-known that a universal enveloping 
algebra of any simple Lie algebra is always a Hopf algebra \cite{4}.
\noindent Let $g$ a simple Lie algebra and $U(g)$ its universal enveloping algebra.\ Then
define the comaltiplication $\bigtriangleup$ ,the counit $\epsilon$ and the antipode
$S$  for $U(g)$  as follows  :
\begin{eqnarray}
\bigtriangleup(x)&=&x \otimes 1  + 1 \otimes x,\\
\epsilon(x)&=&0, \ \ \ \ \ \ \ \ \ \ \ \ \  \ \ \ \ \ \ \ \ \ \ \ \ \ \ \ \ \ \ (x \in   g)    \\
S(x)&=&- x .
\end{eqnarray}
\noindent In particular one can show that the coproduct rule define a Lie algebra
homomorphizm:
\begin{eqnarray}
\bigtriangleup ([x,y])
&=&\bigtriangleup(x)\bigtriangleup(y) -\bigtriangleup(y) \bigtriangleup(x)\\
&=&[\bigtriangleup(x),\bigtriangleup(y)]
\end{eqnarray}
\noindent  For example,\ if one take a Virasoro algebra 
(or also similar infinite dimensional algebras) and feed it into this machinery 
one obtains a Hopf algebra.Although nobody in general taken care of this structure ,the 
purpose of this paper is to confirm this structure in these kind of algebras by using 
the operator product expansion (OPE) language.In this context we note that the universal
enveloping algebra of the underlying symmetry algebra  in the two dimensional (super)conformal
field theory is essentially the same as that of the corresponding universal enveloping algebra of 
any simple Lie algebra. We must also emphasize here that one reason for the importance
of the Hopf algebra is that the Hopf structure of an algebra facilities the construction 
of representations of the sample algebra.

It is known that $W_2$ - algebra 
is equivalent to the Virasoro algebra in the W-algebra framework \cite{5}. The $W_2$-algebra,
involving the modes of a spin-two field $T(z) \equiv \sum_m L_m z^{-m-2}$, is described
by the OPE
\begin{eqnarray}
 T(z) T(w)={{1\over 2} c \over (z-w)^4} +
{{2 T(w)} \over (z-w)^2}+{{\partial{T}(w)} \over {z-w}} \label{yineq}
\end{eqnarray}
\noindent where c is the central charge . Accordingly the Virasoro generators $L_m$'s, which
are the Laurent coefficients of $T(z)$,satisfy the Virasoro algebra
\begin{eqnarray}
[L_n,L_m] = (n-m) L_{n+m}+{c\over{12}} (n^3-n) \delta_{n,-m} , \ \ \ \ \ \ [L_m,c] = 0
\end{eqnarray}

\noindent One must emphasize here that the Virasora algebra is a Lie algebra .
Therefore The Hopf Structure of this algebra is defined by:

\begin{eqnarray}
\bigtriangleup(L_m)&=&L_m \otimes 1 + 1 \otimes L_m,\\
\bigtriangleup(c)&=&c \otimes 1 + 1\otimes c,\\
\epsilon(L_m)&=&0,\\
\epsilon(c)&=&0 ,\\
S(L_m)&=&- L_m,\\
S(c)&=&-c,
\end{eqnarray}

\noindent In this work we will concentrate over conformal field concept, so we will
only recapitulate the above Hopf structure for the energy-momentum tensor $T(z)$
instead of $L_m$ modes  for that reason :

\begin{eqnarray}
\bigtriangleup(T(z))&=&T(z) \otimes 1 + 1 \otimes T(z),\\
\epsilon(T(z))&=&0,\\
S(T(z))&=&- T(z)
\end{eqnarray}
  
\noindent It can be  verified that this comultiplication rule is an algebraic homomrphism
 for $W_2$-algebra (1.6):

\begin{eqnarray}
\bigtriangleup(T(z))\bigtriangleup(T(w))=
  {{\bigtriangleup({1\over 2}c)} \over (z-w)^4} +
{{ \bigtriangleup(2 T(w))} \over (z-w)^2}+{{\bigtriangleup(\partial{T}(w))} \over {z-w}}
\end{eqnarray}

\noindent The organization of this paper is as follows. The next section contains 
the Hopf structure of the Feigin-Fuchs type of free massless  scalar field quantization in the 
commutator and OPE language, respectively.In section 3. the Hopf structure of $W_2$-Algebra is
realized.In section 4.Both the Hopf structure of the Feigin-Fuchs type of free massless fermionic
scalar field quantization and the Hopf structure of $N=1$ superconformal  algebra are realized, 
respectively.

\setcounter{equation}{0}


\section{\bf{The Hopf Structure of Free Massless Bosonic Scalar Fields }}
\vskip 5mm
\noindent A Feigin-Fuchs type of free massless  bosonic scalar field $\varphi (z)$ 
is a single-valued  function on the complex plane and its mode expansion is given by
\begin{eqnarray}
h(z) \ \equiv \ i\,\partial{\varphi}(z)=\sum_{n \in Z} a_n z^{-n-1}.
\end{eqnarray}
Canonical quantization gives the commutator relations
\begin{eqnarray}
[a_m,a_n]\ \equiv\ \ a_m a_n\ - \ a_n a_m\ =\ \kappa m \delta_{_{m+n,0}}\ ,
\end{eqnarray}
where $\kappa$ is a central element commuting with all the modes $\{a_n\}$,
\ $ [a_n,\kappa]=0\ $,and the aim of the central element $\kappa$ is to provide
to Hopf algebra structure for the free field mode algebra (2.2). 
\ This assocoative algebra is a Hopf algebra with
\begin{eqnarray}
\bigtriangleup(a_m)&=&a_m \otimes 1 + 1 \otimes a_m,\\
\bigtriangleup(\kappa)&=&\kappa \otimes 1 + 1 \otimes \kappa,\\
\epsilon(a_m)&=&0,\\
\epsilon(\kappa)&=&0,\\
 S(a_m)&=&- a_m,\\
S(\kappa)&=&-\kappa
\end{eqnarray}
\noindent From the consistency of this Hopf structure with  (2.2),i.e. the
coproduct operation is given by 
\begin{eqnarray}
\bigtriangleup[a_m,a_n] &=& \bigtriangleup(a_m)\bigtriangleup(a_n)-
\bigtriangleup(a_n) \bigtriangleup(a_m)\\
 &=& {1}\otimes [a_m,a_n]+[a_m,a_n]\otimes 1  \\
 &=& \bigtriangleup(\kappa) m \delta_{_{m+n,0}}
\end{eqnarray}
On the other hand,the commutator relations (2.2) are equivalent to the contraction
\begin{eqnarray}
 h(z) h(w) &=&{\kappa \over {(z-w)^2}}+ :h(z) h(w):\ 
\end{eqnarray}
So, we shall now demonstrate that this contraction relation has a Hopf structure in the
OPE language.

\begin{eqnarray}
\bigtriangleup(h(z))&=&h(z) \otimes 1 + 1 \otimes h(z),\\
\epsilon(h(z))&=&0,\\
S(h(z))&=&- h(z)
\end{eqnarray}

\noindent One can  check that  (2.13-15) satisfy the Hopf algebra axioms and
that the defining relation (2.12) is consistent with them. i.e.the coproduct $\bigtriangleup$

\begin{eqnarray}
\bigtriangleup(h(z))\bigtriangleup(h(w))&=& 1\otimes h(z)h(w)+h(z)h(w)\otimes 1\nonumber\\
&&+\ h(z)\otimes h(w)\ + \ h(w)\otimes h(z) \\
&=&{{\bigtriangleup(\kappa)} \over (z-w)^2} +:\bigtriangleup(h(z))\bigtriangleup(h(w)):
\end{eqnarray}

\noindent In the above, it is seen that the last two terms in the (2.16)does not contribute to
 the OPEs  as the singular terms. 
\vskip 5mm

\setcounter{equation}{0}


\section{\bf{The Hopf Structure of $W_2$-Algebra }}

\vskip 5mm

\noindent One can says that the $W_2$-algebra is realized by the Feigin-Fuchs 
type of free massless scalar fields $\{h(z)\}$, of conformal spin-1. 
Let us define a conformal field T(z) having conformal spin-2 as follows:
\begin{eqnarray}
T(z)={1\over 2}: h(z) h(z):
\end{eqnarray}

\noindent Using the contraction (2.12), we want to construct $W_2$-algebra .
So the nontrivial OPE of T(z)  with itself takes the form
\begin{eqnarray}
T(z) T(w)={{{1\over 2} \kappa^2} \over (z-w)^4} +
{{2 \kappa T(w)} \over (z-w)^2}+{{\kappa \partial{T}(w)} \over {z-w}}+:T(z)T(w):
\end{eqnarray}

\noindent For this construction,we used the following  OPEs:
\begin{eqnarray}
T(z) h(w)&=&
{{ \kappa h(w)} \over (z-w)^2}+{{\kappa \partial{h}(w)} \over {z-w}}+:T(z)h(w):,\\ 
h(z) T(w)&=& {{ \kappa h(w)} \over (z-w)^2}+:h(z)T(w):
\end{eqnarray}

\noindent The corresponding Hopf structure is given by

\begin{eqnarray}
\bigtriangleup(T(z))&=&{1\over 2}: \bigtriangleup (h(z))\bigtriangleup (h(z)):\\
&=& 1\otimes T(z)+T(z)\otimes 1 +h(z)\otimes h(z)
\end{eqnarray}

\noindent one can say here that this coproduct is not as in (1.6),but this 
additional term $h(z)\otimes h(z)$ will give us a relation between $\bigtriangleup(c)$
and $\bigtriangleup(\kappa^2)$ in the following calculations.Finally, it can be verified 
that this comultiplication rule is an algebraic homomorphism for
the $W_2$-algebra (3.2).

\begin{eqnarray}
\bigtriangleup(T(z))\bigtriangleup(T(w))\nonumber
&=&1\otimes T(z)T(w)+T(z)T(w)\otimes 1\nonumber\\
&&+\ h(w)\otimes T(z) h(w)+T(z) h(w) \otimes h(w)\nonumber\\
&&+\ h(z)\otimes h(z)T(w)+h(z)T(w)\otimes h(z)\nonumber\\
&&+\ h(z) h(w)\otimes h(z) h(w)
\end{eqnarray}

\noindent by using the explicit operator product expansions (3.2-4) 

\begin{eqnarray}
&=& 1\otimes \left\{
{{{1\over 2} \kappa^2} \over (z-w)^4} +
{{2 \kappa T(w)} \over (z-w)^2}+{{\kappa \partial{T}(w)} \over {z-w}}+:T(z)T(w):\right\}\nonumber\\
&+& \left\{
{{{1\over 2} \kappa^2} \over (z-w)^4} +
{{2 \kappa T(w)} \over (z-w)^2}+{{\kappa \partial{T}(w)} \over {z-w}}+:T(z)T(w):\right\}
\otimes 1\nonumber\\
&+& h(w) \otimes
\left\{{{ \kappa h(w)} \over (z-w)^2}+{{\kappa \partial{h}(w)} \over {z-w}}+:T(z)h(w):\right\}\nonumber\\
&+&\left\{{{ \kappa h(w)} \over (z-w)^2}+{{\kappa \partial{h}(w)} \over {z-w}}+:T(z)h(w):\right\}\otimes h(w)\nonumber\\
&+&\ h(z)\otimes \left\{{{ \kappa h(w)} \over (z-w)^2}+:h(z)T(w):\right\}
\ +\ \left\{{{ \kappa h(w)} \over (z-w)^2}+:h(z)T(w):\right\} \otimes h(z)\nonumber\\
&+&\left\{
{\kappa \over {(z-w)^2}}+ :h(z) h(w):
\right\}
\otimes
\left\{
{\kappa \over {(z-w)^2}}+ :h(z) h(w):
\right\}\nonumber\\
\end{eqnarray}

\noindent and also Taylor expansion for $h(z)$ at $w$
\begin{eqnarray}
&& h(z)\ = \ h(w) + (z-w)\partial h(w)+{1\over 2} (z-w)^2 \partial^2 h(w)+\cdots
\end{eqnarray}

\noindent we obtain
\begin{eqnarray}
\bigtriangleup(T(z))\bigtriangleup(T(w))
 &=&{{\bigtriangleup({1\over 2}\kappa^2)} \over (z-w)^4} +
{{ \bigtriangleup(2 \kappa T(w))} \over (z-w)^2}+
{{ \bigtriangleup( \kappa \partial T(w))} \over z-w}\nonumber\\
&+&:\bigtriangleup(T(z))\bigtriangleup(T(w)):
\end{eqnarray}

\noindent we must emphasize here that the aims of the element $\kappa$ is to provide to Hopf algebra 
structure for the free field algebra (2.12),then the coproduct of central element c for the abstract
$W_2$-algebra (1.6) must be $ \bigtriangleup(c)=c \otimes 1 + 1\otimes c$ as in equation (1.9), but
in the present realization $\tilde c=1$  since there is  one free field and one can says that 
the contribution of the one free field  to the central term is only one ,and then it is seen that the 
central term and  all structure constants of the $W_2$-Algebra  depend only  generator $\kappa$ ,
so there is a relation between the central element $\bigtriangleup(c)$ and $\bigtriangleup(\kappa^2)$ 
in general, but this appears in the present work  as
\begin{eqnarray}
{\bigtriangleup(c)}
&=& \bigtriangleup(\tilde c . \kappa^2)\ 
=\bigtriangleup(\tilde c) \bigtriangleup(\kappa^2)
=\bigtriangleup(\tilde c)\bigtriangleup(\kappa) \bigtriangleup(\kappa)\\
&=&\kappa^2 \otimes 1 + 1\otimes \kappa^2 + 2 \kappa \otimes  \kappa
\end{eqnarray}
\noindent where $ \bigtriangleup(\tilde c=1)=1 \otimes 1 $ . 
Therefore, this relations prevent a doubling of  the central  charge $c$ and $\kappa^2$ .
This points of view  is also valid for N=1 superconformal algebra.
\vskip 5mm

\setcounter{equation}{0}


\section{\bf{The Hopf Structure of Superconformal Algebra }}

\vskip 5mm

\noindent $N=1$ superconformal algebra $\cite{5}$ is generated by a fermionic spin-3/2 chiral
field $G(z)$ and stres-energy tensor $T(z)$ ,which  are satisfy the following OPEs:
\begin{eqnarray}
G(z) G(w)&=&{{2\over 3} c \over (z-w)^3} + {{2 T(w)} \over (z-w)}\\
T(z) G(w)&=&{{3\over 2}G(w) \over (z-w)^2} +{{\partial{G}(w)} \over {z-w}}\\
T(z) T(w)&=&{{1\over 2} c \over (z-w)^4} + {{2 T(w)} \over (z-w)^2}+{{\partial{T}(w)} \over {z-w}}
\end{eqnarray}

\noindent This algebra can be realized  a Feigin-Fuchs Type of free massless scalar field 
$h(z) \ \equiv \ i\,\partial{\varphi}(z)$ (2.1) and a real fermion field

\begin{eqnarray}
\psi(z)=\sum_{n \in Z} \psi_n z^{-n-{1\over 2}}
\end{eqnarray}
\noindent In addition  to the Hopf structure of the Feigin-Fuchs Type of free massless
scalar boson field quantization as in section 2, we have to give the Hopf structure of the 
Feigin-Fuchs Type of free massless  scalar fermion field quantization. So the canonical
quantization gives the following anti-commutator statement,
\begin{eqnarray}
\{\psi_m,\psi_n\}\ \equiv\ \ \psi_m \psi_n + \ \psi_n \psi_m\ =\ \kappa  \delta_{_{m+n,0}}\ 
\end{eqnarray}
\noindent and \ $ [\psi_n,\kappa]=0\ $.\ This assocoative algebra is a Hopf algebra with
\begin{eqnarray}
\bigtriangleup(\psi_m)&=&\psi_m \otimes 1 + 1 \otimes \psi_m, \\
\epsilon(\psi_m)&=&0, \\
S(\psi_m)&=&- \psi_m
\end{eqnarray}
\noindent From the consistency of this Hopf structure with eqn. (4.5),i.e. the
coproduct operation is given by 
\begin{eqnarray}
\bigtriangleup\{\psi_m,\psi_n\}\ &=& \ \bigtriangleup(\psi_m)\bigtriangleup(\psi_n)+
\bigtriangleup(\psi_n) \bigtriangleup(\psi_m)\\
&=& {{1}}\otimes \{\psi_m,\psi_n\}+\{\psi_n,\psi_m\}\otimes 1 \\
&=&\bigtriangleup(\kappa)  \delta_{_{m+n,0}}
\end{eqnarray}
\noindent where we used a parity condition $(a_1\otimes b_1)(a_2\otimes b_2)=-a_1 a_2\otimes b_1 b_2$
(if $b_1 $ and $ a_2 $ are odd). On the other hand,the anti-commutator relations (4.5) are equivalent
 to the contraction statement
\begin{eqnarray}
 \psi(z) \psi(w) ={\kappa \over {z-w}}+ :\psi(z) \psi(w):\ 
\end{eqnarray}
So, we shall now demonstrate that this contraction relation has a Hopf structure in the
OPE language.
\begin{eqnarray}
\bigtriangleup(\psi(z))&=&\psi(z) \otimes 1 + 1 \otimes \psi (z)\\
\epsilon(\psi(z))&=&0\\
S(\psi(z))&=&- \psi(z)
\end{eqnarray}
\noindent One can check that equations (4.13-15) satisfy the Hopf algebra axioms and
that the defining relation equation (4.12) is consistent with them. i.e.the coproduct $\bigtriangleup$
\begin{eqnarray}
\bigtriangleup(\psi(z))\bigtriangleup(\psi(w))&=&1\otimes \psi(z)\psi(w)+\psi(z)\psi(w)\otimes 1\nonumber\\
&&+\psi(z)\otimes\psi(w)+\psi(w)\otimes\psi(z)\\
&=&{{\bigtriangleup(\kappa)} \over {z-w}} +:\bigtriangleup(\psi(z))\bigtriangleup(\psi(w)):
\end{eqnarray}
\noindent In order to realize the Hopf structure of $N=1$ Superconformal algebra which are given as in eqn.(4.1-3).
Let us define a conformal field $G(z)$ having conformal spin-$3\over 2$, as follows :
\begin{eqnarray}
G(z)\ = \ :\psi (z) h(z):   
\end{eqnarray}
\noindent By using the statements (2.12) and (4.12), the OPE of $G(z)$ with
itself  takes the form,
\begin{eqnarray}
G(z) G(w)={{\kappa}^2 \over (z-w)^3} +{{2 \kappa T(w)} \over (z-w)}+:G(z) G(w): 
\end{eqnarray}
\noindent where $T(z)$ is stress-energy tensor,  which is in the form of :
\begin{eqnarray}
T(z)\ = \ {1\over 2}:h(z) h(z):\ + \ {1\over 2} :\psi(z)\partial \psi(z): 
\end{eqnarray}
\noindent and the OPEs with itself  and  also  with $G(z)$are
\begin{eqnarray}
T(z) T(w)&=&{{{1\over 2}\kappa^2} \over (z-w)^4} +
{{2 \kappa T(w)} \over (z-w)^2}+{{\kappa \partial{T}(w)} \over {z-w}}+:T(z) T(w):\\
T(z) G(w)&=&{{3\over 2}\kappa G(w) \over (z-w)^2} +{{\kappa\partial{G}(w)} \over {z-w}}+:T(z) G(w):
\end{eqnarray}
\noindent respectively.For this realization ,we emphasize here that we used the OPEs 
in the equations (3.3-4),and also the following OPEs: 
\begin{eqnarray}
T(z) \psi(w)&=&{{{1\over 2} \kappa \psi(w)} \over (z-w)^2}+{{\kappa \partial{\psi}(w)} \over {z-w}}+:T(z) \psi(w):\\
G(z) h(w)&=&{{\kappa \psi(w)} \over (z-w)^2}+{{\kappa \partial{\psi}(w)} \over {z-w}}+:G(z) h(w):\\
G(z) \psi(w)&=&{{ \kappa h(w)} \over z-w}+:G(z) \psi(w):
\end{eqnarray}
\noindent The Hopf structure of $N=1$ superconformal algebra is given by 
\begin{eqnarray}
\bigtriangleup(G(z))\ &=& :\bigtriangleup(\psi(z)) \bigtriangleup(h(z)): \\
&=& 1\otimes G(z)+G(z)\otimes 1 + h(z)\otimes \psi(z)\ + \psi(z)\otimes  h(z)
\end{eqnarray}
\noindent and
\begin{eqnarray}
\bigtriangleup(T(z))\ &=& \ {1\over 2}:\bigtriangleup(h(z)) \bigtriangleup(h(z)):\ +
 \ {1\over 2}: \bigtriangleup(\psi(z))\bigtriangleup(\partial \psi(z)):\\
&=& 1\otimes T(z)+T(z)\otimes 1 + h(z)\otimes h(z)\nonumber \\
&&+\ \psi(z)\otimes \partial\psi(z)+ \partial \psi(z)\otimes \psi(z)
\end{eqnarray}
Finally, One  can  easily verify that these comultiplication rules are 
an  algebraic homomorphism for the $N=1$ Superconformal algebra (4.1-3).
\begin{eqnarray}
\bigtriangleup(G(z))\bigtriangleup(G(w))&=&{{\bigtriangleup(\kappa}^2) \over (z-w)^3} +
{{\bigtriangleup(2 \kappa T(w))} \over (z-w)}\nonumber\\
&+&:\bigtriangleup(G(z))\bigtriangleup(G(w)): \\
\bigtriangleup(T(z))\bigtriangleup(T(w)) &=&{{\bigtriangleup({1\over 2}\kappa^2)} \over (z-w)^4} +
{{ \bigtriangleup(2 \kappa T(w))} \over (z-w)^2}
+{{\bigtriangleup(\kappa \partial{T}(w))} \over {z-w}}\nonumber\\
&+&:\bigtriangleup(T(z))\bigtriangleup(T(w)):
\end{eqnarray}

\noindent and
\begin{eqnarray}
\bigtriangleup(T(z))\bigtriangleup(G(w)) &=&{{ \bigtriangleup({3\over 2} \kappa G(w))} \over (z-w)^2}
+{{\bigtriangleup(\kappa \partial{G}(w))} \over {z-w}}\nonumber\\
&+&:\bigtriangleup(T(z))\bigtriangleup(G(w)):
\end{eqnarray}
\vskip 3mm


\section{\bf{Conclusions }}
\vskip 3mm
\noindent  In this latter we have  presented that the universal
enveloping algebra of the underlying symmetry algebra  in the two dimensional (super)conformal
field theory is essentially the same as that of the corresponding universal enveloping algebra of 
any simple Lie algebra, with examples only for the $W_2$-algebra and also $N=1$ superconformal
algebra. We will try to extand these studies beyond the $W_2$-algebra and also at least $N=2$
superconformal algebra. The investications in this directions are under study.At  this point this
paper does not have a composed system, but , besides some previous articles \cite{6}                 
a detailed  calculations will be  also given for the connection between the two products in the 
subsequent works which will be the complementary to this one. 


\section{\bf{Acknowledgement}} I would like  to thank M. Chaichian ,M.Arik and E.Hizel
~for their valuable discussions and excellent guidance  throughout this research.

\vskip 5mm


\end{document}